\documentclass[aps,twocolumn,showpacs,groupedaddress]{revtex4}
\usepackage{graphicx}

\begin{document}

% \draft command makes pacs numbers print
%\draft
\title{Effective Hamiltonian approach to adiabatic approximation in open systems}
\author{X. X. Yi$^{1}$,  D. M. Tong$^2$,
L. C. Kwek$^3$, C. H. Oh$^2$}
\affiliation{$^1$Department of Physics, Dalian University of
Technology, Dalian 116024, China\\
$^2$Department of Physics, National University of  Singapore, 10
Kent Ridge Crescent, Singapore 119260\\
$^3$ Department of Natural Sciences, National Institute of
Education, Nanyang Technological University, 1 Nanyang Walk,
Singapore 637616}

\date{\today}

\begin{abstract}
The adiabatic approximation in open systems is formulated through
the effective Hamiltonian approach. By introducing an ancilla, we
embed the open system dynamics into a non-Hermitian quantum
dynamics of a composite system, the adiabatic evolution of the
open system is then defined as the adiabatic dynamics of the
composite system. Validity and invalidity conditions for this
approximation are established and discussed. A High-order
adiabatic approximation for open systems is introduced. As an
example, the adiabatic condition for an open spin-$\frac 1 2$
particle in time-dependent magnetic fields is analyzed.
\end{abstract}

 \maketitle
\section{introduction}
As one of the oldest theorems in quantum mechanics, the adiabatic
theorem\cite{born28} tells us  that if a state is an instantaneous
eigenstate of a sufficiently slowly varying Hamiltonian $H(t)$ at
one time, then it will remain close to that eigenstate up to a
phase factor at later times, while its eigenvalue evolves
continuously.  The adiabatic theorem underlies the adiabatic
approximation scheme, and has potential applications in several
areas of physics such as the Landau-Zener transition in molecular
physics\cite{landau32}, quantum field theory\cite{gell-mann51} and
geometric phase\cite{berry84}. Recently, there has been a growing
interest in the adiabatic approximation in the context of quantum
information, for example, geometric quantum
computation\cite{zanardi99,pachos00,jones00,duan01} and the new
quantum algorithms\cite{farhi00,farhi01}  based on the adiabatic
approximation.

The concept of adiabaticity  has been put forward \cite{sarandy05}
and applied to quantum information\cite{sarandy105} by Sarandy and
Lidar in open systems. In their approach, the open system is
described by a master equation $i\dot{\rho}(t)={\cal L}(t)\rho(t)$
$=[H(t),\rho(t)]+L(t)\rho(t)$, and the adiabatic approximation is
characterized by independently evolving Jordan blocks,  into which
the dynamical superoperator ${\cal L}(t)$ can be decomposed. This
extension for the adiabaticity in open systems is in a systematic
manner, though it is challenging to calculate the superoperator
and determine the Jordan decomposition in practice, especially for
complicated disturbances $L(t)\rho(t)$. The other extension for
the adiabaticity has been presented by Thunstr\"om, {\AA}berg and
Sj\"oqvist for weakly open systems\cite{thunstrom05}. In this
approach, the eigenspace of the system Hamiltonian are primary
instead of instantaneous Jordan blocks. Thus the adiabatic
approximation is  irrespective of the form of $L(t)\rho(t)$. The
main difference between the two approaches is that Sarandy {\it et
al.} focus on the eigenspace of the entire superoperator ${\cal
L}(t)$, while Thunstr\"om {\it et al.} emphasize the decoupling
among the eigenspaces of the system Hamiltonian $H(t)$.

In this paper, we present a new approach to the adiabatic
approximation in open systems by introducing an ancilla to couple
to the quantum system. An effective Hamiltonian which governs the
dynamics of the composite system (quantum system plus ancilla) is
derived from the master equation. We then define the adiabatic
limit of the open system as the regime in which the composite
system evolves adiabatically. The validity and invalidity
conditions for this approximation are also given and discussed. An
extension to high-order adiabatic approximation is presented. As
an example, the adiabatic evolution of a dissipative spin-$\frac 1
2 $ particle driven by time-dependent magnetic fields is analyzed.

The structure of this paper is organized as follows. In Sec. {\rm
II} we embed the open system dynamics into a non-Hermitian quantum
dynamics  by introducing an ancilla. The adiabatic approximation
is defined, and the validity condition is given  in Sec. {\rm
III}. In Sec. {\rm IV}, we present an example to get more insight
on the adiabatic evolution in open systems and discuss the
validity condition. Finally we present our conclusions in Sec.{\rm
V}.

\section{Effective Hamiltonian description for open systems}
We begin with the master equation for the density matrix
$\rho(t)$\cite{gardiner00},
\begin{eqnarray}
i\frac{\partial}{\partial t}\rho(t)&=&[H(t),\rho(t)] -\frac i
2\sum_k
\{L_k^{\dagger}(t)L_k(t)\rho(t)
\nonumber\\
&+&\rho(t)L_k^{\dagger}(t)L_k(t)-2L_k(t)\rho(t)L_k^{\dagger}(t)\}\nonumber\\
&=&[H(t),\rho(t)]+L(t)\rho(t) \nonumber\\
&\equiv&{\cal L}(t)\rho(t), \label{me1}
\end{eqnarray}
where $H(t)$ is a Hermitian Hamiltonian and $L(t)\rho(t)$
describes system-environment couplings and the resulting
irreversibility of decoherence, $L_k(t)$ may be time-dependent
operators describing the system-environment interaction. This
master equation is of the Lindblad form, and thus it guarantees
the conservation of $\mbox {Tr}\rho(t)$ and the positivity of
probabilities. Our aim in this Section is to solve Eq.(\ref{me1})
by introducing an ancilla coupling to the system.  This approach
was first proposed in\cite{yi01} and called effective Hamiltonian
approach. The idea of this approach is as the following. The
density matrix $\rho(t)$ of the open system can be mapped onto a
pure state by introducing an ancilla. The dynamics of the open
system is then described by a Schr\"odinger-like equation with an
effective Hamiltonian that can be derived from the master
equation. In this way the solution of the master equation can be
obtained in terms of the evolution of the composite system by
converting the pure state back to the density matrix. To proceed,
we assume that the dimension of the system Hamiltonian $H(t)$ is
independent of time $t$, so that we may write
$H(t)=\sum_n^N
E_n(t)|E_n(t)\rangle\langle E_n(t)|, $
where $N$ is a constant
representing the dimension of the system. The ancilla labelled by
$A$ is introduced as same as the open system in the sense that its
Hilbert space spanned by $\{|e_m(0)\rangle\}$ has dimension $N$
and remains unchanged in the dynamics. Thus
$\{|E_n(0)\rangle\otimes|e_m(0)\rangle\}$ may be taken as an
orthonormal and complete basis for the composite system. A pure
state for the composite system in the $N^2$-dimension Hilbert
space may be constructed as
\begin{equation}
|\Psi_{\rho}\rangle=\sum_{m,n=1}^N\rho_{mn}(t)|E_m(0)\rangle|e_n(0)\rangle,\label{ee1}
\end{equation}
where $\rho_{mn}(t)$ are density matrix elements of the open
system in the basis $\{|E_m(0)\rangle\}$, i.e.,
$\rho_{mn}(t)=\langle E_m(0)|\rho(t)|E_n(0)\rangle.$ Clearly,
$\langle\Psi_{\rho}(t)|\Psi_{\rho}(t)\rangle={\mbox
Tr}(\rho^2(t))\leq 1$, so this pure bipartite state is generally
not normalized except that the initial state of the open system is
pure and the evolution  is unitary. With these definitions, we now
try to find an effective Hamiltonian ${\cal H}_T(t)$, such that
the bipartite pure state $|\Psi_{\rho}(t)\rangle$ satisfies the
following Schr\"odinger-like equation
\begin{equation}
i\frac{\partial}{\partial t}|\Psi_{\rho}(t)\rangle={\cal
H}_T(t)|\Psi_{\rho}(t)\rangle.\label{se1}
\end{equation}
To simplify the derivation, we write the master equation
Eq.(\ref{me1}) as
\begin{equation}
i\frac{\partial}{\partial t}\rho(t) = {\cal H}(t)\rho(t)
-\rho(t){\cal H}^{\dagger}(t)
 +i\sum_k L_k(t)\rho(t)L_k^{\dagger}(t), \label{me2}
\end{equation}
with ${\cal H}(t)=H(t)-i/2\sum_k L_k^{\dagger}(t)L_k(t).$
Substituting equation Eq.(\ref{ee1}) together with Eq.(\ref{me2})
into Eq.(\ref{se1}), one finds,
\begin{widetext}
\begin{eqnarray}
i\frac{\partial}{\partial
t}|\Psi_{\rho}(t)\rangle&=&\sum_{m,n,p}\langle E_m(0)|{\cal
H}(t)|E_p(0)\rangle\langle
E_p(0)|\rho(t)|E_n(0)\rangle|E_m(0)\rangle|e_n(0)\rangle\nonumber\\
&-& \langle E_m(0)|\rho(t)|E_p(0)\rangle\langle E_p(0)|{\cal
H}^{\dagger}(t)
|E_n(0)\rangle|E_m(0)\rangle|e_n(0)\rangle\nonumber\\
&+&i\sum_k\sum_{m,n,p,q}\langle E_m(0)|
L_k(t)|E_p(0)\rangle\langle E_p(0)| \rho(t)|E_q(0)\rangle\langle
E_q(0)| L_k^{\dagger}(t)|E_n(0)\rangle
|E_m(0)\rangle|e_n(0)\rangle\nonumber\\
&=&\sum_n\left[ {\cal
H}(t)\rho(t)|E_n(0)\rangle|e_n(0)\rangle-{\cal
H}^A(t)\rho(t)|E_n(0)\rangle|e_n(0)\rangle
\right]+i\sum_{k,n}L_k^A(t)L_k(t)\rho(t)|E_n(0)\rangle|e_n(0)
\rangle\nonumber\\
&\equiv&{\cal H}_T(t)|\Psi_{\rho}(t)\rangle,
\end{eqnarray}
\end{widetext}
where ${\cal H}_T(t)$ is defined by
\begin{equation}
{\cal H}_T(t)={\cal H}(t)-{\cal H}^A(t)+i\sum_k L_k^A(t) L_k(t),
\end{equation}
and  referred as  the effective Hamiltonian. Operators ${\cal
H}(t)$ and $L_k(t)$ are for the open system, which take the same
form as in Eq.(\ref{me2}), while ${\cal H}^A(t)$ and $L_k^A(t)$
are operators for the ancilla and defined by
\begin{equation}
\langle e_m(0)|\hat{O}^A|e_n(0)\rangle=\langle E_n(0)|
\hat{O}^{\dagger} |E_m(0)\rangle,
\end{equation}
with $ \hat{O} ={\cal H}(t),$  or $L_k(t).$  The first two terms
in the effective Hamiltonian ${\cal H}_T(t)$ describe the free
evolution of the open system and the ancilla, respectively, while
the third term characterizes couplings between the system and the
ancilla. The other works aiming at setting the passage from pure
to mixed states in an unitary evolution scheme can be found in
Ref.\cite{shadwick01, reznick96,rau02}, but they proceed
differently. In Ref.\cite{shadwick01} the authors have applied the
technique of operator splitting to deal with weakly dissipative
systems. The unitary integrator for the Hamiltonian evolution and
the conventional integrator for the dissipation were combined to
evolve  the open system. Based on the wave operator $\hat{\rho}$
defined by the  density matrix $\rho$,
$\rho=\hat{\rho}\hat{\rho}^{\dagger}$, the unconventional quantum
mechanical formalism was proposed in Ref.\cite{reznick96} to study
the dynamics of open systems. This scheme allows a generalized
unitary evolution between pure and  mixed states, and sheds new
light on the connection between symmetries and conservations laws.
Another work\cite{rau02} deals with open systems by embedding
elements of the density matrix in a higher-dimensional
Liouville-Bloch equation. The dissipation and dephasing in the
open system were included in the non-Hermitian superoperator.
Compared with these schemes, our effective Hamiltonian approach
has the advantage that it is easy to calculate, and as will be
shown  in the next section, the extension for adiabaticity from
closed systems to open systems is straightforward.

\section{The adiabatic approximation in open systems}
In this section, we introduce an adiabatic approximation for open
systems. Conditions for this approximation  are also derived and
discussed. We will restrict our discussions to  systems where the
effective Hamiltonian ${\cal H}_T(t)$ is diagonalizable with
nondegenerate eigenvalues. For further details we refer the reader
to Ref.\cite{sokolov06}, where a general discussion on
non-Hermitian quantum mechanics is presented.

Let us first define the right and left instantaneous eigenstates
of ${\cal H}_T(t)$ by
\begin{eqnarray}
{\cal H}_T(t)|R_m(t)\rangle=\lambda_m(t)|R_m(t)\rangle,\nonumber\\
\langle L_m(t)|{\cal H}_T(t)=\langle
L_m(t)|\lambda_m(t).\label{eigenf1}
\end{eqnarray}
It is easy to show from Eq.(\ref{eigenf1}) that $\langle
L_m(t)|R_n(t)\rangle=\delta_{mn}$ for $\lambda_m\neq\lambda_n$.
Now we are ready to define the adiabatic evolution for open
systems, which is  directly  follow-up  from that for closed
systems. {\it An open system govern by the master equation
Eq.(\ref{me1}) is said to undergo adiabatic evolution if  the
composite system govern by the effective Hamiltonian ${\cal
H}_T(t)$ evolves adiabatically.} In other words, if the effective
Hamiltonian is changed sufficiently slowly, then the composite
system in a given non-degenerate eigenstate of the initial
effective Hamiltonian ${\cal H}_T(0)$ evolves into the
corresponding eigenstate of the instantaneous Hamiltonian ${\cal
H}_T(t)$, without making any population transitions. This leads to
the definition of adiabatic evolution in the open system which is
the corresponding adiabatic dynamics in the composite system. This
definition is a straightforward extension  of the idea of
adiabatic evolution for open systems, and it will be shown below
that the condition for this to occur backs to the conventional
adiabatic evolution when the system is a closed system. Let us now
derive the validity conditions for open system adiabatic
evolution. To this end, we expand $|\Psi_{\rho}(t)\rangle$ for an
time $t$  in the instantaneous right eigenstates of ${\cal
H}_T(t)$ as
\begin{equation}
|\Psi_{\rho}(t)\rangle=\sum_m c_m(t)e^{-i\int_0^t \lambda_m(\tau)
d\tau }|R_m(t)\rangle.\label{expan1}
\end{equation}
 Substituting
$|\Psi_{\rho}(t)\rangle$ in    into the Schr\"odinger-like
equation Eq.(\ref{se1}), one obtains
\begin{equation}
\dot{c}_m(t)+\langle L_m(t)|\dot{R}_m(t)\rangle
c_m(t)=c_{off}(t),\label{dife1}
\end{equation}
where
\begin{equation} c_{off}(t)=-\sum_{n\neq m}c_n(t)\langle
L_m(t)|\dot{R}_n(t)\rangle
e^{-i\int_0^t(\lambda_n(\tau)-\lambda_m(\tau))d\tau}.
\end{equation}
Formal integration of Eq.(\ref{dife1}) yields,
\begin{eqnarray}
c_m(t)&=&c_m(0)-\int_0^t\langle L_m(\tau)|\dot{R}_m(\tau)\rangle
c_m(\tau)d\tau\nonumber\\
&+&\int_0^t c_{off}(\tau)d\tau.
\end{eqnarray}
In accordance with the definition of adiabaticity in open systems,
the adiabatic regime is obtained when $C_{off}(t)=\int_0^t
c_{off}(\tau)d\tau$ is negligible. This condition ensures that the
mixing of coefficients $c_m(t)$ corresponding to distinct
eigenvalues $\lambda_m(t)$  is absent, which in turn guarantees
that the change in ${\cal H}_T(t)$ is sufficient slow. The latter
claim can be shown by rewriting the adiabatic condition
$C_{off}(t) \rightarrow 0$ as
\begin{eqnarray}
\Gamma_{mn}(t)&=&\left|\frac{\langle L_n(t)|\partial {\cal
H}_T(t)/\partial t
|R_m(t)\rangle}{(\lambda_m(t)-\lambda_n(t))^2}\right|\nonumber\\
&=&\left|\frac{\langle L_n(t)|\dot{R}_m(t)\rangle}{
\lambda_m(t)-\lambda_n(t)} \right|\ll 1.\label{adiacon1}
\end{eqnarray}
In terms of the density matrix $\rho(t)$, $\langle
L_n(t)|\dot{R}_m(t)\rangle$ can be expressed as $\langle
L_n(t)|\dot{R}_m(t)\rangle={\mbox Tr }(\rho_n(t)\dot{\rho}_m(t)),$
where the elements of $\rho_m(t)$ are  defined as
$[\rho_m(t)]_{\alpha\beta}=$ $\langle e_{\beta}(0)|\langle
E_{\alpha}(0)|R_m(t)\rangle $, i.e., $\rho_m(t)$ is the density
matrix corresponding to the $m$-th right eigenstate of the
effective Hamiltonian. Then the adiabatic condition in this case
becomes $|{\mbox Tr} (\rho_n(t)\dot{\rho}_m(t))|\ll
|\lambda_m(t)-\lambda_n(t)|.$ This relation implies that the
transition rate from one path $\rho_n(t)$ to the other path
$\rho_m(t)$ is negligible in the adiabatic evolution.
\begin{figure}
\includegraphics*[width=0.8\columnwidth,
height=0.4\columnwidth]{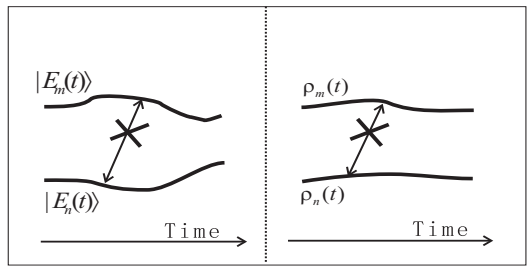} \caption{In closed systems,
the transition rate among distinct instantaneous eigenstates $\{
|E_m(t)\rangle \}$of a varying Hamiltonian is zero in adiabatic
evolution(left), whereas it is zero among distinct density
matrices $\{ \rho_m(t) \}$ in open systems (right). The distinct
density matrices $\rho_m(t)$ can be obtained by mapping the
instantaneous eigenstates of the effective Hamiltonian back to the
open system.} \label{fig2}
\end{figure}
From the other aspect, the adiabatic evolution for open systems
indicates that $|\langle L_m(t)|\Psi_{\rho}(t)\rangle|$ remains
constant in the dynamics for any $m$(degenerate levels are
excluded). This results in ${\mbox Tr } (\rho_m(t)\rho(t))=$
constant, showing again that transition rate among distinct paths
$\rho_m(t)$ is zero in the adiabatic dynamics. A comparison of the
adiabatic evolution in closed systems with that in open systems
was sketched in figure \ref{fig2}. For a closed system,
$L(t)\rho(t)=0$, yielding ${\cal H}_T(t)=H(t)-H^A(t)$, where
$H(t)$ is the system Hamiltonian given in Eq.(\ref{me1}), $H^A(t)$
is the counterpart of $H(t)$ for the ancilla. Note that $H^A(t)$
takes the same form as $H(t)$, except that it operates on the
ancilla. Then the adiabatic condition in this case backs to
\begin{equation}
\left|\frac{\langle
E_m(t)|\dot{E}_n(t)\rangle}{E_m(t)-E_n(t)}\right|\ll 1.
\end{equation}
This is the well known adiabatic condition for closed
system\cite{marzlin04}. In other words, the adiabatic condition
defined here for open systems back to the adiabatic condition for
closed systems when $L(t)\rho(t)=0,$ showing the consistency of
the definition  for open systems.

The definition of adiabatic evolution can be easily generalized to
high-order adiabatic approximation. To this end, we take the
instantaneous eigenspace of the system Hamiltonian $H(t)$ as the
primary Hilbert space. As will be shown, this choice has the
advantage that the formalism would return straightforwardly  to
Thunstr\"om's results for weakly open systems. Defining
\begin{equation}
U(t)=\sum_{n}^N|E_n(0)\rangle\langle E_n(t)|,
\end{equation}
and
\begin{equation}
\tilde{\rho}(t)=U(t)\rho(t)U^{\dagger}(t),
\end{equation}
one gets a master equation for $\tilde{\rho}(t)$ from
Eq.(\ref{me1})\cite{thunstrom05},
\begin{eqnarray}
i\frac{\partial}{\partial
t}\tilde{\rho}(t)&=&[\tilde{H}(t),\tilde{\rho}(t)]+[Z(t),\tilde{\rho}(t)]\nonumber\\
&-&\frac i 2\sum_k
\{\tilde{L}_k^{\dagger}(t)\tilde{L}_k(t)\tilde{\rho}(t)
+\tilde{\rho}(t)\tilde{L}_k^{\dagger}(t)\tilde{L}_k(t)\nonumber\\
&-&2\tilde{L}_k(t) \tilde{\rho}(t)\tilde{L}_k^{\dagger}(t)\},
\label{me3}
\end{eqnarray}
where
$$\tilde{H}(t)=\sum_n E_n(t)|E_n(0)\rangle\langle E_n(0)|,$$
$$Z(t)=i\dot{U}(t)U^{\dagger}(t),\ \ \tilde{L}_k(t)=U(t)L_k(t)U^{\dagger}(t).$$
The effective Hamiltonian corresponding to this master equation
can be derived by the same procedure and given by
\begin{equation}
\tilde{{\cal H}}_T(t)=\tilde{{\cal H}}(t)-\tilde{{\cal
H}}^A(t)+i\sum_k \tilde{L}_k^A(t)\tilde{L}_k(t),
\end{equation}
where $\tilde{{\cal H}}(t)=\tilde{H}(t)+Z(t)-\frac i 2\sum_k
\tilde{L}_k^{\dagger}(t)\tilde{L}_k(t).$ The high-order adiabatic
evolution is then defined as the adiabatic evolution of the
composite system with the effective Hamiltonian $\tilde{{\cal
H}}_T(t)$. The reason of referring the adiabatic approximation
with $\tilde{{\cal H}}_T(t)$ as the hight-order adiabatic
approximation is that $Z(t)$ contains off-diagonal terms
$\sum_{m,n}|E_m(0)\rangle\langle E_m(t)|\dot{E}_n(t)\rangle\langle
E_n(0)|$, $m\neq n$. In the adiabatic approximation defined with
${\cal H}_T(t),$ these terms have been ignored when the open
system approaches to a closed system. However, they  could not be
ignored in the latter definition.  In fact, the adiabatic
approximation defined with $\tilde{{\cal H}}_T(t)$ is for
$\tilde{\rho}(t)$ in the master equation (\ref{me3}), which is
different from $\rho(t)$ in Eq.(\ref{me1}), and this definition is
of consistency with the high-order approximation in closed
systems. This point can be found by the same analysis presented
above for the definition with ${\cal H}_T(t)$.
\section{example: the adiabatic evolution of an open spin-$\frac 1
2 $ particle in time-dependent magnetic fields}
In this section,
we present an example to get more insight of the adiabatic
evolution in open systems. The example consists of a dissipative
spin-$\frac 1 2 $ particle driven by a time-dependent magnetic
field. The master equation govern the dynamics of such a system
can be written as
\begin{equation}
\dot{\rho}=-i[H(t),\rho]+\frac{\kappa}{2}\{
2\sigma_-\rho\sigma_+-\rho\sigma_+\sigma_--\sigma_+\sigma_-\rho\},
\end{equation}
where $H(t)=\mu \vec{B}(t)\cdot \vec{\sigma}$ denotes the system
Hamiltonian, $\kappa$ stands for the spontaneous emission rate.
$\sigma_i, i=z,+,- $ are Pauli matrices, and
$\sigma_z=|e\rangle\langle e|-|g\rangle\langle g|$ ($|e\rangle$,
denotes the  state of spin-up, and $|g\rangle$ spin-down). Suppose
$\vec{B}(t)=B_0(\sin\theta\cos\phi,\sin\theta\sin\phi,
\cos\theta),$ the effective Hamiltonian corresponding to $\rho(t)$
reads,
\begin{equation}
{\cal H}_T(t)= {\cal H} (t)-  {\cal H}^A (t)+i\kappa
\sigma_-\tau_-,
\end{equation}
where ${\cal H} (t)=H(t)-\frac i 2|e\rangle\langle e|,$   ${\cal
H}^A (t)=H^A(t)+\frac i 2|e\rangle_A\langle e|,$ and
$\tau_-=|g\rangle_A\langle e|$ represents the Pauli matrix of the
ancilla. In a subspace spanned by
$\{|gg\rangle,|ge\rangle,|eg\rangle, |ee\rangle\}$, the effective
Hamiltonian can be written as (in units of $\mu B_0$),
\begin{widetext}
\begin{equation}
{\cal H}_T(t)=\left( \matrix{ 0 &  -\sin\theta e^{-i\phi} &
\sin\theta e^{i\phi} & i\gamma \cr
   -\sin\theta e^{i\phi} &
  -2\cos\theta-0.5i\gamma & 0 &  \sin\theta e^{i\phi} \cr
 \sin\theta e^{-i\phi} & 0 & 2\cos\theta-0.5i\gamma &  -\sin\theta
 e^{-i\phi}\cr
 0 &  \sin\theta e^{-i\phi} &  -\sin\theta e^{i\phi} & -i\gamma
   } \right),\label{eff2}
\end{equation}
\end{widetext}
where $\gamma=\kappa/\mu B_0$, and the first(second) letter in the
basis $|ab\rangle (a,b=e,g)$ stands for states of the
system(ancilla). The eigenvalues $\lambda_j$ (in units of $\mu
B_0$) of ${\cal H}_T(t)$ are given by
\begin{eqnarray}
 \lambda_1&=&0,\nonumber\\
(\lambda_j&+&0.5i\gamma)^3+0.5i\gamma(\lambda_j+0.5i\gamma)^2-4(\lambda_j+0.5i\gamma)\nonumber\\
&=&2i\gamma^2\cos^2\theta, \ \  ( j=2,3,4),
\end{eqnarray}
and the corresponding right eigenstates
\begin{equation}
|R_j\rangle=\frac{1}{\sqrt{M_j}} \left( \matrix{ a_j   \cr
   b_j\cr
 c_j\cr
d_j } \right),
\end{equation}
as well as  the left eigenstates
\begin{equation}
\langle L_j|=\frac{1}{\sqrt{M_j}} \left( \matrix{ A_j,   &
   B_j,&C_j,& D_j } \right).
\end{equation}
Here
\begin{eqnarray}
M_j&=&A_ja_j+B_jb_j+C_jc_j+D_jd_j,\nonumber\\
a_1&=&1+\frac{4\cos^2\theta+0.25\gamma^2}{\sin^2\theta},\nonumber\\
b_1&=&-\frac{2\cos\theta-0.5\gamma i e^{i\phi}}{\sin\theta},\nonumber\\
c_1&=&-\frac{2\cos\theta+0.5\gamma i e^{-i\phi}}{\sin\theta},\nonumber\\
d_1&=&1;\nonumber\\
A_1&=&D_1=1, B_1=C_1=0.\nonumber\\
\end{eqnarray}
\begin{eqnarray}
\mbox {For} \ \ j&=&2,3,4, \nonumber\\
a_j&=&-(2\cos\theta+0.5i\gamma+\lambda_j),\nonumber \\
b_j&=&2\sin\theta e^{i\phi},\nonumber\\
c_j&=&\frac{2\sin\theta(2\cos\theta+0.5i\gamma+\lambda_j)e^{-i\phi}}{2\cos\theta-0.5i\gamma-\lambda_j},\nonumber\\
d_j&=&-a_j,\nonumber\\
A_j&=&-(2\cos\theta-0.5i\gamma+\lambda_j),\nonumber \\
D_j&=&\frac{i\gamma-\lambda_j}{i\gamma+\lambda_j}A_j,\nonumber\\
B_j&=&\frac{\sin\theta
e^{-i\phi}(D_j-A_j)}{2\cos\theta+0.5i\gamma+\lambda_j},\nonumber\\
C_j&=&\frac{\sin\theta
e^{i\phi}(D_j-A_j)}{2\cos\theta-0.5i\gamma-\lambda_j},\nonumber\\
\end{eqnarray}
To simplify the discussion, we assume that $\phi=\omega t$ with a
constant $\omega$, and $\theta$  remains unchanged in the
dynamics. With this assumption, it is readily to show that all
eigenvalues of ${\cal H}_T(t)$ are time-independent, so $M_j$ are
constant. To show the dependence of the adiabatic condition on the
spontaneous emission rate $\gamma$ and $\omega$, we define the
following function with $max$ taken over all $m$ and
$n$\cite{note2},
\begin{equation}
\Gamma(\gamma,\omega)=max \left\{ \left| \frac{\langle
L_n(t)|\dot{R}_m(t)\rangle}{\lambda_m-\lambda_n}\right| \right\},\
\ m,n=1,2,3,4,
\end{equation}
which characterizes  the violation of the adiabatic evolution.
 \begin{figure}
\includegraphics*[width=0.8\columnwidth,
height=0.6\columnwidth]{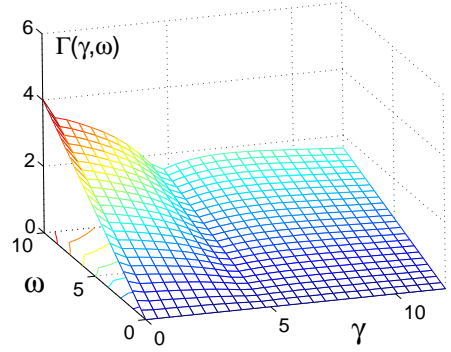} \caption{(Color online)An
illustration of $\Gamma=\Gamma(\gamma,\omega)$ as a function of
$\omega$ and $\gamma$.  $\omega$ was chosen in units of $\mu B_0$,
and this plot is for $\theta=\frac 1 4\pi$. } \label{fig1}
\end{figure}
The numerical results of $\Gamma(\gamma,\omega)$ versus $\gamma$
and $\omega$ were illustrated in figure 1. The rotating frequency
$\omega$ of the driven field was plotted in units of $\mu B_0$,
where $B_0$ is the modulus of the driving field. Figure 1 shows us
that $\Gamma(\gamma,\omega)$ depends on $\omega$ linearly with a
fixed $\gamma$, while $\Gamma(\gamma,\omega)$ decreases  with
$\gamma$ increasing. This can be understood as the following.
Without the driving field $\vec{B}(t)$, the dependance of
$|R_m(t)\rangle$ on $\gamma$ behaves like $e^{-\gamma t}$, while
$\lambda_m$ depends on $\gamma$ linearly. Therefore,
$\Gamma(\gamma,\omega)$ decays with $\gamma$ increasing and tends
to zero when $\gamma\rightarrow \infty.$
\section{summary and discussion}
We have presented an adiabatic approximation scheme for open
systems. In contrast with the conventional adiabatic
approximation, the adiabatic approximation for open systems have
been defined as the adiabatic dynamics of a composite system,
which consists of the system and an ancilla. Our
effective-Hamiltonian based definition of adiabaticity  retains
the conventional  adiabatic approximation in the ideal case of
closed systems, hence it is of consistency. The definition of
adiabaticity in open systems has been extended to a high-order
adiabatic approximation, which was defined in accordance with an
master equation for the rotated density matrix. The validity and
invalidity condition of adiabticity has been derived and
discussed.  The violation of adiabatic evolution has been
demonstrated  by an field-driven dissipative spin-$\frac 1 2 $
particle. The other applications and demonstrations such as
geometric phase in open systems and the effect of decoherence on
quantum adiabatic computing will be addressed elsewhere.

\ \ \\
This work was supported by EYTP of M.O.E,  NSF of China (10305002
and 60578014), and the NUS Research Grant No.
R-144-000-071-305.\\


\begin{references}
\bibitem{born28}
{M. Born and V. Fock}, {Z. Phys.} {\bf 51},  165  (1928).

\bibitem{landau32}
{L. D. Landau}, Zeitschrift {\bf 2},  46  (1932); {C. Zener},
Proc. R. Soc. London Ser. A {\bf 137},  696 (1932).


\bibitem{gell-mann51}
{M. Gell-Mann and F. Low}, Phys. Rev. {\bf 84},  350  (1951).

\bibitem{berry84} M. V. Berry, Proc. R. Soc. London A {\bf 392},
45(1984).

\bibitem{zanardi99}
{P. Zanardi and M. Rasetti}, Phys. Lett. A {\bf 264},  94  (1999).


\bibitem{jones00}
{J. A. Jones, V. Vedral, A. Ekert, and G. Castagnoli}, Nature
(London) {\bf 403},  869 (2000).

\bibitem{pachos00}
{J. Pachos and S. Chountasis}, Phys. Rev. A {\bf 62},  052318
(2000).

\bibitem{duan01}
{L.-M. Duan, J. I. Cirac, and P. Zoller}, Science {\bf 292},  1695
(2001).

\bibitem{farhi00}
{E. Farhi, J. Goldstone, S. Gutmann, and M. Sipser}, e-print
quant-ph/0001106.

\bibitem{farhi01}
{E. Farhi, J. Goldstone, S. Gutmann, J. Lapan, A. Lundgren, and D.
Preda}, Science {\bf 292},  472  (2001).

\bibitem{sarandy05} M. S. Sarandy and D. A. Lidar, Phys. Rev. A
{\bf 71}, 012331(2005).

\bibitem{sarandy105} M. S. Sarandy and D. A. Lidar, Phys. Rev.
Lett. {\bf 95}, 250503(2005).

\bibitem{thunstrom05} P. Thunstr\"om, J. {\AA}berg, and E.
Sj\"oqvist, Phys. Rev. A {\bf 72}, 022328(2005).

\bibitem{gardiner00} C. W. Gardinar and P. Zoller, Quantum noise
(Springer, Berlin, 2000).

\bibitem{yi01} X. X. Yi and S. X. Yu, J. Opt. B: Quantum
Semiclass. {\bf 3}, 372(2001).

\bibitem{shadwick01} B. A. Shadwick and W. F. Buell, J. Phys. A
{\bf 34}, 4771(2001).

\bibitem{reznick96} B. Reznick, Phys. Rev. Lett. {\bf 76},
1192(1996).

\bibitem{rau02} A. R. P. Rau and R. A. Wendell, Phys. Rev. Lett.
{\bf 89}, 220405(2002).

\bibitem{sokolov06} A. V. Sokolov, A. A. Andrianov, and F.
Cannata, e-print: quant-ph/0602207.

\bibitem{marzlin04}For recent progresses in this direction, please
read, K. P. Marzlin, B. C. Sanders, Phys. Rev. Lett. {\bf 93},
160408(2004); D. M. Tong {\it et al.}, Phys. Rev. Lett. {\bf 95}
110407(2005). As shown, this condition may not be sufficient for
some special systems, however, it is powerful in most cases. So
does Eq.(\ref{adiacon1}).

\bibitem{note2}Degenerate levels are excluded.


\end{references}
\end{document}